\documentclass[twocolumn,showpacs,preprintnumbers,amsmath,amssymb]{revtex4}
\usepackage{graphics, lscape, afterpage, longtable, bm}
\begin{document}
\title{Poisson-Bracket Approach to the Dynamics of Bent-Core Molecules}
\author{William Kung}\email{w-kung@northwestern.edu}
\affiliation{Department of Materials Science and Engineering, Northwestern University, Evanston, Illinois 60208}
\author{M. Cristina Marchetti}
\affiliation{Department of Physics, Syracuse University, Syracuse, New York 13244}
\date{\today}

\begin{abstract}
We generalize our previous work on the phase stability and hydrodynamic of polar liquid crystals possessing local uniaxial $C_{\infty v}$-symmetry to biaxial systems exhibiting local $C_{2v}$-symmetry.  Our work is motivated by the recently discovered examples of thermotropic biaxial nematic liquid crystals comprising bent-core mesogens, whose molecular structure is characterized by a non-polar body axis $({\bf{n}})$ as well as a polar axis $({\bf{p}})$ along the bisector of the bent mesogenic core which is coincident with a large, transverse dipole moment.  The free energy for this system differs from that of biaxial nematic liquid crystals in that it contains terms violating the ${\bf{p}}\rightarrow -{\bf{p}}$ symmetry.  We show that, in spite of a general splay instability associated with these parity-odd terms, a uniform polarized biaxial state can be stable in a range of parameters.  We then derive the hydrodynamic equations of the system, via the Poisson-bracket formalism, in the polarized state and comment on the structure of the corresponding linear hydrodynamic modes.  In our Poisson-bracket derivation,  we also compute the flow-alignment parameters along the three symmetry axes in terms of microscopic parameters associated with the molecular geometry of the constituent biaxial mesogens.
\end{abstract}


\maketitle

\section{Introduction}

In the past few decades, the field of liquid crystallography has benefited from a tremendous progress in the increasing sophistication of technology and experimental designs.  The ability to manipulate fine microscopic details of constituent molecules, such as geometrical shape, electric and magnetic moments, as well as chemical affinity, has enabled the creation of many  interesting liquid-crystalline phases of ever increasing complexity, both in terms of their characterization and properties.  Since Freiser's seminal theoretical work~\cite{Freiser1970} predicting that compounds composed of biaxial molecules should in principle engender a biaxial nematic phase, and the experimental discovery of the lyotropic biaxial nematic phase a decade later by Yu and Saupe~\cite{Yu1980}, the experimental realization of a thermotropic biaxial nematic phase has long been the holy grail of liquid crystallography.  Aside from the potential technological applications, biaxial nematic systems also hold theoretical interest in that they provide the simplest example of an ordered medium with a non-Abelian fundamental homotopy group~\cite{Mermin1979}.  In the ensuing decades, there were several claims~\cite{Malthete86, Chandrasekhar88, Praefcke90} to the discovery of systems displaying biaxial nematic phase as confirmed by optical techniques.  However, under the scrutiny of the more reliable method of NMR spectroscopy~\cite{Luckhurst01}, these prior claims were proven to be false hopes.  In the mean time, novel engineering attempts were made involving more complex systems, such as binary mixtures of hard rods and plates, that exhibit a macroscopic phase with two symmetry axes~\cite{Raton2002}, where each species in these binary mixtures orientate in different directions that are orthogonal to each other.  Finally, in 2004 unequivocal evidence materialized demonstrating the existence of the thermotropic biaxial nematic phase based on the techniques of NMR spectroscopy~\cite{Madsen04, Luckhurst04} and X-ray scattering~\cite{Acharya2004}.  In these experiments, the elusive biaxial nematic phase was realized by systems composed of bent-core mesogenic molecules, which had first garnered attention for the discovery by Niori {\it{et al.}}~\cite{Niori1996} because they exhibited unusual smectic phases with ferroelectric and antiferroelectric properties.    It has been postulated~\cite{Madsen04, Luckhurst04} that strong intermolecular associations that originate from the large electric dipole moment pointing towards the center of these ``V"-shaped mesogens, rather than their pure geometry, should play a crucial role in the stability of the biaxial nematic phase.  One strong evidence supporting this claim comes from the experimentally observed apex angle, formed by the two arms of each mesogenic molecule, of 140 degrees which is far from the predicted theoretical optimum value of 109.4 degrees~\cite{Teixeira1998, Luckhurst2001}.  Thus, it is an important consideration that one of the two molecular symmetry axes of these bent-core mesogens, along which the direction of the dipole moment lies,  is polar.

Such general issues on the geometry of mesogenic molecules and the stability of the resulting macroscopic phases have been investigated by a few numerical studies~\cite{Teixeira1998, Luckhurst2001, Bates2005, Wilson2006}.   Specifically, it has been found that when the angle between the arms of these bent-core molecules becomes the tetrahedral (which has generally not been observed in mesogenic systems exhibiting the biaxial nematic phase), the isotropic phase undergoes a continuous phase transition into the biaxial nematic phase~\cite{Teixeira1998, Luckhurst2001}.  The effect of the anisotropy in the length of the two arms as well as the angle between them has been investigated in a more recent Monte Carlo simulation~\cite{Bates2005}.  Analytically, a few other theoretical issues have also been looked at,  including static properties such as stability of phases~\cite{Straley1974, Boccara1977, Longa2005, Vissenberg1997, Mettout2005, Mettout2002}, elasticity and viscosity~\cite{Longa1998a,Kapanowski1997,Longa1998b,Fialkowski1998}, and critical behavior~\cite{Mukherjee1998}. The dynamical properties~\cite{Brand1982,Pleiner1985,Jacobsen1981,Galerne1986,Brand1981, Liu1981} of biaxial nematic systems have also been considered.  For the most part, many of the aforementioned theoretical works have been mainly focused on biaxial systems exhibiting the macroscopically nonpolar $D_{2h}$-symmetry~\cite{Freiser1970, Straley1974, Boccara1977,Longa2005,Vissenberg1997, Mettout2005,Brand1982,Pleiner1985,Jacobsen1981,Galerne1986,Brand1981}, in which both of the symmetry axes are non-polar.
\begin{figure}[t]
\vspace{2mm}
\hspace{23mm}
\includegraphics{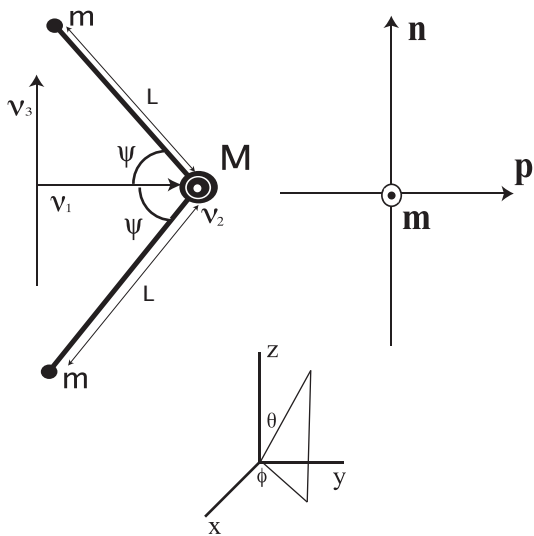}
\vspace{2mm}
\caption{A schematic representation of the bent-core mesogen~\cite{Lubensky2002}, with orthonormal molecular axes \{\boldmath$\hat{\nu}_1, {\hat{\nu}}_2, \hat{\nu}_3$\unboldmath\}.  The molecular dipole points along \boldmath$\hat{\nu}_1$\unboldmath, while the molecular body axis lies along \boldmath$\hat{\nu}_3$\unboldmath;\; \boldmath$\hat{\nu}_2$\unboldmath\, points into the page.   Macroscopically, we define the orthonormal triad $\{{\bf{n, p, m}}\}$ with the polar and nematic director fields along ${\bf{p}}$ and ${\bf{n}}$, respectively. }
\label{mnp}
\end{figure}

In this paper, we focus on the stability and dynamics of the polar biaxial phase $(C_{2v})$ that preserves the symmetry of the molecular polar structure of these bent-core mesogens.  The alternative scenario of macroscopic polarity in a biaxial nematic phase, wherein two symmetry axes are polar, has been considered in~\cite{Mettout2002}.  To fully characterize this {\it{bipolar}} biaxial phase,  the spatial orientation of the molecules must be specified by the full set of Euler angles, thereby generalizing the corresponding distribution function from the spherical harmonics $\left\{Y^l_m\right\}$ to the Wigner $D$-functions $\left\{D^L_{m,m'}\right\}$~\cite{Dray1986}, which form a complete set of orthogonal functions with respect to integration over $SO(3)$.  The necessary order parameters, of their bipolar biaxial nematic phase, that exhibits local $C_s$-symmetry, then reduce naturally to the components of two vectors.  In more general instances, the most direct consequence of biaxiality is the generalization of the alignment tensor to contain the usual Maier-Saupe order-parameter as well as a new {\it{biaxial}} order-parameter.  For concreteness, we define our coordinate systems as shown in Fig.~\ref{mnp}.  We assume that the dipole moment and the body axis of a bent-core mesogen point along its molecular axes \boldmath$\hat{\nu}_1$\unboldmath and \boldmath$\hat{\nu}_3$\unboldmath, respectively.  The third unit vector identifying the molecular coordinate system is \boldmath$\hat{\nu}_2$\unboldmath $=$ \boldmath$\hat{\nu}_3$\unboldmath $\times$ \boldmath$\hat{\nu}_1$\unboldmath, such that
$\sum_a{\hat{\nu}}^{\alpha}_{a,i}{\hat{\nu}}^{\alpha}_{a,j}=\delta_{ij}$.

For  general systems exhibiting biaxial symmetry,  achiral ordered phases are fully characterized by both the second-rank biaxial alignment tensor $Q_{ij}$ and the alignment vector ${\bf{P}}$.  These in turn can be written as
%
\begin{eqnarray}
\label{Qb}
Q_{ij}({\bf{r}})
&=&S({\bf{r}})\left[n_i({\bf{r}})n_j({\bf{r}})-\frac13\delta_{ij}\right]\nonumber\\
&&+T({\bf{r}})\left[p_i({\bf{r}})p_j({\bf{r}})-m_i({\bf{r}})m_j({\bf{r}})\right]\;,\\
\nonumber\\
%
{\bf{P(r)}}
&=&S_p({\bf{r}}){\bf{p(r)}}+S_n({\bf{r}}){\bf{n(r)}}+S_m({\bf{r}}){\bf{m(r)}}\;,
\label{Pb}
\end{eqnarray}
where $S=\langle P_2(\cos\theta)\rangle$ and $T=\langle \sin^2\theta\cos 2\phi\rangle$, with $\theta$ and $\phi$ denoting the polar and azimuthal angles, respectively.   The unit vectors $\left\{ {\bf{p}}, {\bf{m}}, {\bf{n}}\right\}$ are an orthonormal triad.  In~\cite{Lubensky2002}, Lubensky and Radzihovsky introduced a third-order tensor parameter for the bent-core molecules in order to fully characterize the chiral phases also observed in these systems.  Since we do not consider such chiral phases in our work, it suffices to consider only the alignment tensor and vector as defined in Eqs. (\ref{Qb}) and (\ref{Pb}).

In the polar biaxial phase where the symmetry of the macroscopic phase coincides with that of the constituent molecules $(C_{2v})$, we assume that, on average, the non-polar body axis of the mesogens points along ${\bf{n(r)}}$, while the polar axis induced by the microscopic dipole moments lies along the unit vector ${\bf{p(r)}}$.  Thus, $S_n=S_m=0$.  Furthermore, ${\bf{n(r)}}$ and ${\bf{p(r)}}$ are always orthogonal by construction and ${\bf{m}}={\bf{n\times p}}$.  The vector ${\bf{m}}$ is therefore not an independent quantity.  The biaxial alignment tensor can be rewritten solely in terms of ${\bf{p}}$ and ${\bf{n}}$ as
\begin{eqnarray}
Q_{ij}&=&\left(S+T\right)\left(n_in_j-\frac13\delta_{ij}\right)+2T\left(p_ip_j-\frac13\delta_{ij}\right)\;.\nonumber\\
\end{eqnarray}
The scalar quantity $S$ is the usual Maier-Saupe order parameter that describes alignment along ${\bf{n(r)}}$, the direction corresponding to the maximum eigenvalue of $n_iQ_{ij}n_j=\frac{2}{3}S$.  The {\it{biaxial}} order parameter $T$ describes ordering of the mesogens in the ${\bf{m-p}}$ plane.  The direction along ${\bf{p}}$ corresponds to the second-largest eigenvalue of $p_iQ_{ij}p_j=-\frac13S+T$.  The scalar field $S_p$ encodes information on the degree of local alignment of the dipoles~\cite{Kung2006}.  The characterization of biaxal nematic phases with different local symmetries would generally require different sets of order parameters.

A comprehensive classification of possible phases for bent-core molecules has been given by Lubensky and Radzihovsky~\cite{Lubensky2002} and also by Mettout~\cite{Mettout2005}.
With the identification of the proper order parameters, group-theoretic arguments and Landau mean-field theory can be used to catalogue the universality classes of these phase transitions.  It has been found in \cite{Lubensky2002} that transitions to the polar biaxial phase with $C_{2v}$-symmetry $\left\{S_p\neq 0, S\neq 0, T\neq 0\right\}$ are allowed from various higher-symmetry phases, including the polar unaxial $(C_{\infty v})$ phase $\left\{S_p\neq 0, S\neq 0, T= 0\right\}$ and the uniaxial nematic $(D_{\infty h})$ phase $\left\{S_p= 0, S\neq 0, T= 0\right\}$, as well as the nonpolar biaxial nematic $(D_{2h})$ phase $\left\{S_p= 0, S\neq 0, T\neq 0\right\}$.  Assuming that the nematic director field aligns along ${\bf{n}}$ and the polar director field along ${\bf{p}}$, one can envision the transition to the polar biaxial $(C_{2v})$-phase from the polar uniaxial phase  by the development of biaxial order in the plane perpendicular to ${\bf{p}}$, or similarly, from the uniaxial nematic phase via the establishment of a vector order along ${\bf{p}}$.  Rotational invariance in the plane normal to the director ${\bf{n}}$ implies that both of these transitions will be in the $XY$-universality class~\cite{Lubensky2002, PathriaText}.  On the other hand, the transition to the polar biaxial $(C_{2v})$ phase from the nonpolar biaxial nematic $(D_{2h})$ phase is characterized by the development of vector order along one of the three twofold symmetries already specified in the biaxial nematic phase and thus renders the transition a member of the Ising universality class with well-known critical properties~\cite{Lubensky2002, PathriaText}.

In what follows, we will investigate the stability and the dynamics of the polar biaxial phase $(C_{2v})$, in the linear hydrodynamics regime, thereby generalizing the results of our prior work~\cite{Kung2006} on the uniaxial case.  We derive the hydrodynamic equations from a microscopic model using the Poisson-bracket formalism and obtain expression for the flow-alignment parameters.  The structure of the hydrodynamic equations in the biaxial case differs qualitatively from the unaxial case, since biaxiality breaks completely the rotational group $SO(3)$.  This symmetry breaking generates a total of three Goldstone modes, in contrast to only two found in the uniaxial case due to the residual $SO(2)$ group.  The structure of the corresponding hydrodynamic modes is also more complicated due to the discrete orthorhombic symmetry in the biaxial mesogens, which is a lower symmetry group than the cylindrical symmetry found in uniaxial mesogens.  The existence of polarity  again implies additional parity-odd terms in the free energy that would otherwise be forbidden in the nonpolar case.  The most important of such terms couple density fluctuations and polarization splay and render the uniform polarized biaxial phase unstable in a range of parameters.

Our derivation starts in section II where we introduce a microscopic definition of the various order parameters.  In section III, we consider the stability of the polarized phase from the viewpoint of the free energy.  In particular, we show that polarity can be compatible with orientational order in liquid phases. We then obtain the hydrodynamic equations for the biaxial polarized phase in section IV.  We consider these equations in the linear hydrodynamic regime and comment on the structure of the resulting hydrodynamic modes in section V.  Section VI concludes our paper.

\section{Definitions of Microscopic Field Variables}

In our previous work~\cite{Kung2006}, we studied the hydrodynamics of polar liquid crystals with local $C_{\infty v}$-symmetry.  We applied the Poisson-bracket formalism~\cite{Volovik1980a, Volovik1980b} to derive the hydrodynamic equations for the conserved variables---mass and momentum densities---as well as for the polar director field.  We obtained a microscopic expression for the flow alignment parameter $\lambda$~\cite{Brand1982, Forster1974, Carlsson1992, Saupe1981} and showed that the hydrodynamic equation for the polar director field have the same form as its nematic counterpart, albeit with a different $\lambda$~\cite{Stark2003, Forster1971, Orsay1969, Lubensky1970}.  As mentioned in the Introduction, we have introduced the polarization ${\bf{P}}$ as the order parameter characterizing the polar liquid phase in uniaxial systems.  In the case of a biaxial phase in which one of the two director fields is polar, we will need to employ the alignment tensor $Q_{ij}$ in conjunction to the alignment or  polarization vector ${\bf{P}}$ in order to fully characterize the polar phase.

We consider a system composed of $N$ identical biaxial mesogens, indexed by $\alpha$.    Based on the three-atom model in~\cite{Lubensky2002}, we assume that each molecule consists of three atoms, treated as point particles, along the two arms of the boomerang, as shown in Fig. (\ref{mnp}).  Each molecule is described by a molecular coordinate system $\{$\boldmath$\hat{\nu}_1, \hat{\nu}_2, \hat{\nu}_3 $\unboldmath $\}$ as defined in Fig. (\ref{mnp}).  The position of the center of mass of the $\alpha$-th molecule is given by

\begin{eqnarray}
{\bf{r}}^{\alpha}&=&\left(\frac{M}{2m+M}L^{\alpha}\cos\psi^{\alpha}\right){\bm{{\hat{\nu}}}}^{\alpha}_1\;.
\label{Rcm}
\end{eqnarray}
The atoms of mass $m$ have charges $q$, and the central atom of mass $M$ possesses charge $Q$.  We assume that molecules are neutral,
\begin{equation}
2q+Q=0\;,
\end{equation}
but possess a non-vanishing dipole moment ${\bf{d}}^{\alpha}$, given by
\begin{eqnarray}
{\bf{d}}^{\alpha}=QL^{\alpha}\cos\psi^{\alpha}\,{\bm{ {\hat{\nu}}}}^{\alpha}_1\;,
\end{eqnarray}
where ${\bm{{\hat{\nu}}}}^{\alpha}_1$ is the unit vector along the polar molecular axis (Fig.~\ref{mnp}).  We note that while the molecular dipolar interactions are physically important in stabilizing the observed biaxial liquid-crystalline phase, the asymmetry in the molecular shapes alone also gives rise to polarity in the system.     The alignment vector ${\bf{P}}$ can then be defined as the coarse-grained sum of the microscopic {\it{polar}} axes,
\begin{equation}
{\bf{P(r)}}=\frac{m_0}{\rho{\bf{(r)}}}\left[\sum_{\alpha}{\bm{{\hat{\nu}}}}^{\alpha}_1\delta({\bf{r-r^{\alpha}}})\right]_c\;,
\label{Pdef}
\end{equation}
where $m_0=M+2m$, $\rho({\bf{r}})$ is the local mass density, and can be expressed in terms of a unit vector ${\bf{p}}$, as given in Eq. (\ref{Pb}) with, $S_n=S_m=0$.  Analogously, the alignment tensor is defined in terms of a coarse-grained sum of the traceless, second molecular mass-moment tensor ${\hat{R}}_{ij}^{\alpha}$:
\begin{eqnarray}
Q_{ij}=\frac{1}{R{\bf{(r)}}}\left[\sum_{\alpha}{\hat{R}}^{\alpha}_{ij}\delta(\bf{r}-\bf{r}^{\alpha})\right]_c\;,
\label{Qmicro}
\end{eqnarray}
where
\begin{eqnarray}
{\hat{R}}_{ij}^{\alpha}&=&\sum_{\mu}m_{\mu}\left[\Delta r^{\alpha\mu}_i\Delta r^{\alpha\mu}_j-\frac13\delta_{ij}\left(\Delta {\bf{r}}^{\alpha\mu}\right)^2\right]\;.
\end{eqnarray}
Here $\mu$ labels the atoms within each molecule and $\Delta {\bf{r}}^{\alpha\mu}={\bf{r}}^{\alpha\mu}-\Delta {\bf{r}}^{\alpha}$ is the position of each atom relative to the center of mass.  The molecular mass distribution tensor ${\hat{R}}_{ij}$ can also be written with respect to the basis of the orthonormal molecular coordinate system $\{$\boldmath $ {\hat{\nu}}_a^{\alpha}$\unboldmath$\}$ as
\begin{eqnarray}
{\hat{R}}_{ij}^{\alpha}&=&\sum_a\,R^{\alpha}_aQ^{\alpha}_{a, ij},
\label{RR}
\end{eqnarray}
with
\begin{eqnarray}
Q^{\alpha}_{a,ij}&=&{\hat{\nu}}^{\alpha}_{a,i}{\hat{\nu}}^{\alpha}_{a,j}-\frac13\delta_{ij}\;\,\,\,\,\,\,\,\,\,\,\,\,\,\,\,\,\,\,\,\,\,\,\,\,a = 1, 2, 3\;,
\label{Qdef}
\end{eqnarray}
and
\begin{eqnarray}
R^{\alpha}_1&=&-R^{\alpha}_2=-\frac{mM}{2m+M}\left(L^{\alpha}\right)^2\cos^2\psi^{\alpha},\\
\nonumber\\
R^{\alpha}_3&=&2m\left(L^{\alpha}\right)^2\sin^2\psi^{\alpha}-\frac{mM}{2m+M}\left(L^{\alpha}\right)^2\cos^2\psi^{\alpha}.\nonumber\\
\end{eqnarray}
Finally, the scalar field $R({\bf{r}})$ is given by
\begin{eqnarray}
R({\bf{r}})&=&\sum_{\alpha, a}R^{\alpha}_a\delta({\bf{r}}-{\bf{r}}^{\alpha})\;,\\
&=&\sum_{\alpha}R^{\alpha}_3\delta({\bf{r}}-{\bf{r}}^{\alpha})\;.
\label{RR}
\end{eqnarray}
Equivalently, we can express the coefficients $\left\{R^{\alpha}_a\right\}$ in terms of the molecular moment-of-inertia tensor
\begin{eqnarray}
I^{\alpha}_{ij}&=&\sum_{\mu}m^{\mu}\left[\left(\Delta {\bf{r}}^{\alpha\mu}\right)^2\delta_{ij}-\Delta r^{\alpha\mu}_i\Delta r^{\alpha\mu}_j\right]\;,
\end{eqnarray}
whose components along the principal axes are related to the second molecular mass-moment coefficients as follows:
\begin{eqnarray}
R^{\alpha}_1&=&-\frac{I^{\alpha}_{33}}{2}=-R^{\alpha}_2\;,\\
R^{\alpha}_3&=&I^{\alpha}_{11}-\frac{I^{\alpha}_{33}}{2}=I^{\alpha}_{22}-\frac{3I^{\alpha}_{33}}{2}\;.
\end{eqnarray}
We can then write down, in a straightforward manner, the trace $I^{\alpha}$
\begin{eqnarray}
I^{\alpha}&\equiv&{\rm{Tr}}\,I^{\alpha}_{ij}=I^{\alpha}_{11}+I^{\alpha}_{22}+I^{\alpha}_{33}\nonumber\\
&=&4m\left(L^{\alpha}\right)^2\sin^2\psi^{\alpha}+\frac{4mM}{2m+M}\left(L^{\alpha}\right)^2\cos^2\psi^{\alpha}\;,\nonumber\\
\label{trace}
\end{eqnarray}
as well as the anisotropy $\Delta I^{\alpha}$
\begin{eqnarray}
\Delta I^{\alpha}&\equiv&I^{\alpha}_{11}-\frac12I^{\alpha}_{33}\;,\nonumber\\
&=&2m\left(L^{\alpha}\right)^2\sin^2\psi^{\alpha}-\frac{mM}{2m+M}\left(L^{\alpha}\right)^2\cos^2\psi^{\alpha}\;.\nonumber\\
\label{AI}
\end{eqnarray}
Subsequently, we neglect fluctuations in the molecular shape and thus assume that $R^{\alpha}_a$ is thus independent of $\alpha$.  We can then write
\begin{eqnarray}
R({\bf{r}})&=&\frac{\Delta I}{m_0}\rho({\bf{r}})\;.
\end{eqnarray}
In the limit of $\psi\rightarrow \pi/2$, its molecules become needle-like, and we recover the expected ratio of $I/\Delta I=2$~\cite{Stark2003, Footnote}.

The biaxial alignment tensor in Eq. (\ref{Qmicro}) can thus be identified with the macroscopic expression in Eq. (\ref{Qb}) expressed in terms of orthogonal unit vectors ${\bf{m}}$, ${\bf{n}}$, and ${\bf{p}}$.

\section{Free energy and stability of the polarized phase}

Before proceeding with the derivation of the hydrodynamic
equations for the polarized state, it is instructive to discuss
the coarse-grained free energy of a polar, biaxial fluid.
Compared with the polar unaxial phase, the free energy becomes
much more complex due to the overall lower symmetry of the biaxial
phase.  We will again divide the free energy into a parity-even
contribution that respects the symmetries of ${\bf{n}}\rightarrow
-{\bf{n}}$ and ${\bf{p}}\rightarrow -{\bf{p}}$ and a parity-odd
contribution that arises from the polar nature of ${\bf{p}}$,
$F=F^{\rm{even}}+F^{\rm{odd}}$.  The number of distinct elastic
constants is dictated by the discrete symmetries of the system.
Starting with a general elastic free energy with terms quadratic
in the gradients, {\it{i.e.}} terms of the forms
$\left(\nabla_in_j\right)\left(\nabla_ln_m\right)$,
$\left(\nabla_ip_j\right)\left(\nabla_lp_m\right)$, and
$\left(\nabla_in_j\right)\left(\nabla_lp_m\right)$, and minding
symmetry considerations, we find that there are in total fifteen
independent terms in the elastic free energy of biaxial nematics.
Using the anholonomity (Mermin-Ho) relations for the three broken
rotational symmetries~\cite{Mermin1976, Kamien2002}, which first
appeared in the context of superfluid phases of
helium-3~\cite{Pleiner1983a, Kopf1997, Graham1974, Kopf1999,
Kovalevsky2003}, three of the fifteen terms can be shown to be
surface contributions~\cite{Brand1982}.  The Mermin-Ho relations,
which reflect the consequence of the non-commutative nature of
finite rotations, can be expressed in terms of their
locally-defined infinitesimal generators
$\delta\bm{\Theta}=\left(\delta\Theta_1, \delta\Theta_2,
\delta\Theta_3\right)=\left({\bf{m}}\cdot\delta{\bf{n}},
-\left({\bf{p}}\cdot\delta{\bf{n}}\right),
{\bf{p}}\cdot\delta{\bf{m}}\right)$:
\begin{eqnarray}
\left(\delta_1\delta_2-\delta_2\delta_1\right)\bm{\Theta}&=&\left(\delta_1\bm{\Theta}\right)\times\left(\delta_2\bm{\Theta}\right)\;,
\end{eqnarray}
where  $\delta_1$ and $\delta_2$ represent any first-order
differential operators, such as $\partial_t$ or $\bm{\nabla}$.
For example,  to the first-order approximation of an infinitesimal
variation of the director ${\bf{n}}$, defined with respect to the
free-energy functional, $d\nabla_jn_i\neq\nabla_jdn_i$.

The remaining twelve terms in the elastic free energy are bulk
contributions. The bulk elastic terms and the mass terms that are
invariant under parity are given by
\begin{widetext}
\begin{eqnarray}
F^{\rm{even}}&=&\frac12\int\,d^3r\left\{C_1\left(\frac{\delta\rho}{\rho_0}\right)^2+C_2\left(\bm{\nabla}\rho\right)^2+K_{n1}\left(\bm{\nabla}\cdot{\bf{n}}\right)^2+K_{n2}\left({\bf{n}}\cdot\bm{\nabla}\times{\bf{n}}\right)^2+K_{n3}\left({\bf{n}}\times\bm{\nabla}\times{\bf{n}}\right)^2+K_{p1}\left(\bm{\nabla}\cdot{\bf{p}}\right)^2\right.\nonumber\\
&&+K_{p2}\left({\bf{p}}\cdot\bm{\nabla}\times{\bf{p}}\right)^2+ +K_{p3}\left({\bf{p}}\times\bm{\nabla}\times{\bf{p}}\right)^2+K_{m1}\left[\bm{\nabla}\cdot\left({\bf{p\times n}}\right)\right]^2 +K_{m2}\left[\left({\bf{p\times n}}\right)\cdot\bm{\nabla}\times\left({\bf{p\times n}}\right)\right]^2\nonumber\\
&&\left.+K_{m3}\left[\left({\bf{p\times n}}\right)\times\bm{\nabla}\times\left({\bf{p\times n}}\right)\right]^2+K_{mp}\left[\left({\bf{p\times n}}\right)\cdot\left(\bm{\nabla}\times{\bf{p}}\right)\right]^2+K_{pn}\left[{\bf{p}}\cdot\left(\bm{\nabla}\times{\bf{n}}\right)\right]^2\right.\nonumber\\
&&\left.+K_{nm}\left[{\bf{n}}\cdot\left(\bm{\nabla}\times\left({\bf{p\times n}}\right)\right)\right]^2\right\}\;,
\label{Feven}
\end{eqnarray}
\end{widetext}
where $\delta\rho=\rho-\rho_0$ is the local fluctuation in
density. The elastic free energy can be written down in many
equivalent forms.  The first nine elastic constants, from $K_{n1}$
up to $K_{m3}$, have the simple interpretation of describing the
bulk deformations (splay, twist, and bend) along the three
symmetry axes, respectively.  The last three terms have no simple
analogues in uniaxial elasticity.  Theoretical calculations based
on the microscopic approach~\cite{Longa1998a, Kapanowski1997,
Longa1998b} have been done to determine the magnitudes of these
elastic constants from first principles, as the results strongly
suggested a match to what one would intuitively expect that these
``cross"-terms should take on smaller magnitudes than the other
nine elastic constants.

The existence of polarity allows for terms that break the
${\bf{p}}\rightarrow -{\bf{p}}$ in the elastic free energy. The
lowest-order contribution to the parity-odd free energy has the
form:
\begin{widetext}
\begin{eqnarray}
F^{\rm{odd}}&=&\int\,d^3r\left\{B\frac{\delta\rho}{\rho_0} \left(\bm{\nabla}\cdot{\bf{p}}\right)+D_p\left(\bm{\nabla}\cdot{\bf{p}}\right)\left({\bm{\nabla}}\cdot{\bf{n}}\right)^2+O\left(\bm{\nabla}^4\right)\right\}\;.\nonumber\\
\label{Fodd}
\end{eqnarray}
\end{widetext}
By construction, Eq. (\ref{Fodd}) involves only odd orders of the polar director ${\bf{p}}$.  As in the unaxial case, terms $\sim\bm{\nabla}\cdot{\bf{p(r)}}$ give rise to surface terms that would favor a splay distortion of the mesogens,  though the instability can be suppressed assuming appropriate surface stabilization.  In addition to possible surface instability, it is apparent from the form of the new parity-odd $B$-term that a ${\it{bulk}}$ instability can also develop.  This intrinsic bulk instability to splay deformations has been well known in the uniaxial case~\cite{Kung2006}.  We can understand qualitatively the existence of this bulk instability by rewriting the terms in the free energy involving density fluctuations and splay deformations of the director field:
\begin{widetext}
\begin{eqnarray}
\int \frac{d^3r}{2}
\Big\{C_1\Big(\frac{\delta\rho}{\rho_0}\Big)^2+K_{p1}(\bm{\nabla}\cdot\delta{\bf{
p}})^2+2B\frac{\delta\rho}{\rho_0}\left(\bm{\nabla}\cdot\delta{\bf
p}\right)\Big\}= \int \frac{d^3r}{2}
\Big\{\frac{C_1}{\rho_0^2}\big[\delta\rho+\frac{B\rho_0}{C_1}\big(\bm{\nabla}\cdot\delta{\bf
p}\big)\big]^2
+\big[K_{p1}-\frac{B^2}{C_1}\big](\bm\nabla\cdot\delta{\bf
p})^2\Big\}\;,\nonumber\\
\label{Fpp}
\end{eqnarray}
\end{widetext}
where $\delta{\bf{p}}={\bf{p}}-{\bf{p}}_0$ is the fluctuation of
the polar director field about its equilibrium orientation.  Thus,
the existence of polarity along a symmetry axis in the biaxial
phase, which allows for new parity-odd terms in the free energy,
yields an overall downward renormalization of the splay elastic
constant $K_{p1}$.  We emphasize again that the molecular asymmetry of the
bent-core molecules is sufficient to induce the aforementioned
polarity.  Though in actual experimental realization of the
biaxial phase $(S_p=0, S\neq 0, T\neq 0)$, the existence of a
strong molecular dipole moment and the resulting dipolar
interactions are conjectured to be the essential contributing
factor to the stability of this phase.  Our discussion remains
valid as long as polarity exists in the system, regardless of the
nature of its origin.

While completing the square in the free energy expression of Eq.
(\ref{Fpp}) demonstrates that the splay instability is a general
{\it{bulk}} property of the polarized state, this simple argument does not capture the anisotropic nature of
the instability that will become apparent below from the analysis of the hydrodynamic modes of a polar biaxial system to be carried out in section V.  The anisotropy
of the splay instability is also fully apparent  form a purely static calculation of the 
density-density correlation function of the system. Such a calculation must be carried out  enforcing the
orthonormality constraint that ${\bf{n\cdot p}}=0$ which is not incorporated in the free energy.

The free-energy as written in Eq. (\ref{Fpp}) indicates  that splay fluctuations become energetically favorable and destabilize the ordered state when $\vert B\vert > \sqrt{K_{p1}C_1}$.  The instability is suppressed if the fluid is incompressible since the renormalization of the Frank constant $K_{p1}$ vanishes when $C_1\rightarrow\infty$.  The stable ground state for $\vert B\vert > \sqrt{K_{p1}C_1}$ is expected to be characterized by director configurations that are spatially inhomogenous in a direction that is perpendicular to the polar, broken symmetry director ${\bf{p(r)}}$, with associated spatial structures in the density~\cite{Blankschtein1985}.

We can examine the possible relevance of the splay instability to experimental systems by performing numerical estimates of the various elastic constants.  We consider a fluid of mesogens of linear size $l$ interact via a short-range repulsive interaction of strength $U$.  The mesogens forming the biaxial phase possess a permanent dipole moment of magnitude $d$ that yields a dipolar coupling of strength $U_D\sim k_cd^2/r^3$ between two mesogenic molecules at distance $r$, where $k_c$ is the Coulomb's constant.  The splay Frank coefficient has dimensions of energy/length and can be estimated as $K_{p1}\sim U/l$, and the bulk compressional modulus $C_1$ has dimensions of energy density.  Its magnitude is controlled by the repulsive part of the interaction and thus can be estimated as $C_1\sim U/l^3$.  The elastic constant $B$ is controlled by interactions that favor alignment of the mesogenic molecules and would otherwise vanish if the molecules are nonpolar.  Thus, we estimate $B\sim U_D/l^2$, where $U_D$ is the strength of the dipolar interaction.  In our estimate, splay instability will occur when $U_D\sim U$.  Since $r\sim\rho_0^{-1/3}$, we take $\rho_0\sim 10^{-27}$m and $d\sim 4$ Debye~\cite{Madsen04, Acharya2004} and estimate $U_D\sim 0.01\,{\rm{eV}}$.  While there are no reported measurement of the elastic constants for thermotropic biaxial liquid crystals, theoretical analysis in~\cite{Palangana2003} has established a mapping between the elastic constants of a biaxial sample and of an uniaxial sample under suitablly chosen geometry and boundary conditions, in which the biaxial elastic constants should be comparable in magnitude to their unaxial counterparts.   Thus, we estimate that $U\sim 0.1{\rm{eV}}$.  Consequently, we see that a stable polar biaxial phase with local $C_{2v}$-symmetry is, in principle, possible with appropriate surface stabilization.

\section{Hydrodynamic equations in the Polarized Phase}

The hydrodynamic equations for uniaxial liquid crystals have been rigorously derived, on the basis of the Poisson-bracket formalism in both the nematic phase $(D_{\infty h})$ and the polarized $(C_{\infty v})$ phase~\cite{Kung2006, Stark2003}.  It has been shown that the dynamical equation for the director field retains the same form in both phases, and the effect of polarity manifests itself only in a different microscopic expression of the flow-alignment parameter, $\lambda$~\cite{Kung2006}.  For systems exhibiting macroscopic phases of biaxial symmetry, the identification of all three spatial directions (versus just one in the unaxial case) leads to three distinct flow-alignment parameters.  Any system that breaks the same continuous symmetry should obey hydrodynamic equations with identical structures~\cite{Martin1972}.  Thus, continuous symmetry of a system dictates the nature of its hydrodynamic variables as well as the characterization of propagating and diffusive modes, in contrast to discrete symmetries which determine issues such as the coupling of modes.   In the isotropic phase with full rotational and translational invariance, if we restrict our attention to isothermal processes, the hydrodynamic fields consist only of the four conserved variables:  number density and the three components the momentum density (we do not consider energy).  In the biaxial phase, we have three additional hydrodynamic fields due to broken rotational symmetry.  Therefore, the total set of hydrodynamic field variables, $\left\{\Phi_a({\bf{r}},t)\right\}$, in the biaxial case, consists of the number density $\rho({\bf{r}})$, momentum density ${\bf{g(r)}}$, nematic director field ${\bf{n(r)}}$, and polar director field ${\bf{p(r)}}$.  We note that the director fields ${\bf{n(r)}}$ and ${\bf{p(r)}}$ are not independent but satisfy the orthogonality constraint ${\bf{n\cdot p}}=0$.  The conserved and broken-symemtry variables evolve in time according to~\cite{Stark2003, Kung2006}
\begin{eqnarray}
\frac{\partial\Phi_a({\bf{r}},t)}{\partial t}&=&V_a-\Gamma_{ab}\frac{\delta F}{\delta\Phi_b({\bf{r}})}\;,
\label{Evo}
\end{eqnarray}
where $\Gamma_{ab}$ is the symmetric dissipative tensor, $F$ is the total free energy shown in Eqs (\ref{Feven}) and (\ref{Fodd}), and $V_a$ is the reactive term or nondissipative velocity given by
\begin{eqnarray}
V_a({\bf{r}})&=&-\int d^3r'\,\left\{\Phi_a({\bf{r}}), \Phi_b({\bf{r'}})\right\}\frac{\partial\mathcal{H}}{\partial\Phi_b({\bf{r'}})}\;,
\label{V}
\end{eqnarray}
where the Poisson bracket between the field variables in Eq. (\ref{V}) is defined as
\begin{eqnarray}
\{\Phi_a({\bf{r}}),\Phi_b({\bf{r'}})\}&=&\sum_{\alpha,\mu}\sum_i\left(\frac{\partial \Phi_a({\bf{r}})}{\partial p^{\alpha\mu}_{ i}}\,\frac{\partial \Phi_b({\bf{r'}})}{\partial r^{\alpha\mu}_{ i}}\right.\nonumber\\
&&-\left.\frac{\partial \Phi_a({\bf{r}})}{\partial r^{\alpha\mu}_{
i}}\, \frac{\partial \Phi_b({\bf{r'}})}{\partial p^{\alpha\mu}_{
i}}\right)\;.
\label{PBdefinition}
\end{eqnarray}
and can be computed as coarse-grained expressions of suitably defined microscopic variables.  In the following, we will present a summary of the necessary ingredients in the derivation of hydrodynamics via the Poisson-bracket method.  The detailed evaluation of the various Poisson brackets can be found in the Appendix.

\subsection{Poisson-bracket Formalism}

Using Eq. (\ref{PBdefinition}), it is
straightforward to evaluate the Poisson-bracket  relations between
the various hydrodynamic fields.  Our goal is to obtain equations
describing the dynamics at long wavelengths.  We will therefore
only keep terms of lower order in the gradients of the
hydrodynamic fields and expand the $\delta$-function as
\begin{eqnarray}
\delta({\bf{r}}-{\bf{r^{\alpha\mu}}})&=&\delta({\bf{r}}-{\bf{r^{\alpha}}}-\Delta{\bf{r^{\alpha\mu}}})\nonumber\\
&=&\delta({\bf{r}}-{\bf{r^{\alpha}}})-\Delta{{r^{\alpha\mu}_{ k}}}\nabla_k\delta({\bf{r}}-{\bf{r^{\alpha}}})\nonumber\\
&&+O(\nabla^2)\;.
\label{delta}
\end{eqnarray}
where $\Delta{\bf{r^{\alpha\mu}}}={\bf{r}}^{\alpha\mu}-{\bf{r^{\alpha}}}$.  In addition to the microscopic definitions of polarization and alignment tensor in Eqs. (\ref{Pdef}) and (\ref{Qdef}), we have the microscopic mass density, ${\hat{\rho}}({\bf{r}})$ and momentum
density, ${\hat{{\bf{g}}}}({\bf{r}})$, defined in the usual way as
\begin{eqnarray}
{\hat{\rho}}({\bf{r}})&=&\sum_{\alpha,\mu}m^{\mu}\delta({\bf{r}}-{\bf{r^{\alpha\mu}}})\;,\\
{\bf{\hat{g}(r)}}&=&\sum_{\alpha,\mu}{\bf{p^{\alpha\mu}}}\,\delta({\bf{r}}-{\bf{r^{\alpha\mu}}})\;,
\end{eqnarray}
where ${\bf p^{\alpha\mu}}$ is the momentum of the $\mu$-th atom
on the $\alpha$-th molecule. The macroscopic mean variables
describing the dynamics of equilibrium fluctuations in the system
are obtained from the microscopic ones after coarse-graining,
$\rho({\bf{r}})=\left[{\hat{\rho}}({\bf{r}})\right]_c$,
${\bf{g}}({\bf{r}})=\left[{\hat{\bf{g}}}({\bf{r}})\right]_c$,
as described, for instance, in Ref. \cite{Stark2003}.  The required non-vanishing Poisson-brackets are:
\begin{eqnarray}
\label{pg}
\{\rho({\bf{r}}),g_i({\bf{r'}})\}&=&\nabla_i\delta({\bf{r-r'}})\rho({\bf{r'}})\;,\\
\label{gp}
\{g_i({\bf{r}}),g_j({\bf{r'}})\}&=& g_i({\bf
r'})\nabla_j\delta({\bf{r-r'}})\nonumber\\
&&-\nabla'_i\left[\delta({\bf{r-r'}}) g_j({\bf r'})\right]\;,\\
\label{Qg}
\{Q_{ij}({\bf{r}}),g_k({\bf{r'}})\}&=&\left[\nabla_kQ_{ij}({\bf{r}})\right]\delta({\bf{r-r'}})\nonumber\\
&&-\lambda_{ijkl}({\bf{r}})\nabla_l\delta({\bf{r-r'}})\;,
\end{eqnarray}
where
\begin{eqnarray}
\lambda_{ijkl}&=&\delta_{ik}Q_{jl}+\delta_{jk}Q_{il}-\frac23\delta_{ij}Q_{kl}-\frac23\delta_{kl}Q_{ij}\nonumber\\
&&+\frac{I}{6\Delta I}\left(\delta_{ik}\delta_{jl}+\delta_{il}\delta_{jk}-\frac23\delta_{ij}\delta_{kl}\right)\nonumber\\
&&+O(Q^2_{ij})\;,
\label{lambdaijkl}
\end{eqnarray}
in which we average both the trace and anisotropy over the system and thus sum over the index $\alpha$ in Eqs. (\ref{trace}) and (\ref{AI}).  The Poisson-bracket relation shown in Eqs. (\ref{Qg}) and (\ref{lambdaijkl}) has been previously obtained by Stark and Lubensky in~\cite{Stark2003} for uniaxial nematic liquid crystals.  In their expression of $\lambda_{ijkl}$, they also obtained a second-order term in $Q_{ij}$ by the inclusion of the dynamics of the amplitude variable of $R_{ij}$, which ultimately did not contribute to the hydrodynamical equation for the nematic director.  For biaxial liquid crystals, both Eqs. (\ref{Qg}) and (\ref{lambdaijkl}) retain the same form when we use our molecular moment-of-inertia expression [Eq. (\ref{RR})] in the Poisson-bracket derivation.

The director fields themselves have no microscopic definition in terms
of the canonical coordinates and momenta of the individual atoms.
Their Poisson-brackets must therefore be obtained indirectly from
those of the alignment tensor $Q_{ij}$. Using the definition of
the polar and nonpolar directors given in Eq.~(\ref{Qb}) and the chain rule of
derivatives we obtain, respectively,
\begin{eqnarray}
\{p_i({\bf{r}}),g_j({\bf{r'}})\}&=&\frac{1}{2T}\left[\delta^{T_p}_{ik}(\bm{r})-\left(\frac{S+T}{S-T}\right)n_i(\bm{r})n_k(\bm{r})\right]\nonumber\\
&&\times\left\{Q_{kl}({\bf{r}}),g_j({\bf{r'}})\right\}p_l({\bf{r}})\;. \label{pg3}
\end{eqnarray}
\begin{eqnarray}
\{n_i({\bf{r}}),g_j({\bf{r'}})\}&=&\frac{1}{S} \left[\delta^{T_n}_{ik}(\bm{r})+\frac{T}{S-T}p_i(\bm{r})p_k(\bm{r})\right.\nonumber\\
&&\left.-\frac{T}{S+T}m_i(\bm{r})m_k(\bm{r})\right]\nonumber\\
&&\times\left\{Q_{kl}({\bf{r}}),g_j({\bf{r'}})\right\}n_l({\bf{r}})\;. \label{ng3}
\end{eqnarray}
where
\begin{eqnarray}
\label{Tp}
\delta^{T_p}_{ij}(\bm{r})&=&\delta_{ij}(\bm{r})-p_i(\bm{r})p_j(\bm{r})\;,\\
\delta^{T_n}_{ij}(\bm{r})&=&\delta_{ij}(\bm{r})-n_i(\bm{r})n_j(\bm{r})
\label{Tn}
\end{eqnarray}
are the projection operators onto components perpendicular to ${\bf{p}}$ and ${\bf{n}}$, respectively.  Upon insertion of Eq. (\ref{Qg}), we obtain
\begin{eqnarray}
\{p_i({\bf{r}}),g_j({\bf{r'}})\}
&=&\left[\nabla_jp_i(\bm{r})\right]\delta\left({\bf{r}}-{\bf{r}}'\right)\nonumber\\
&&-\frac{1}{2T}\left(\delta^{T_p}_{ik}(\bm{r})-\frac{S+T}{S-T}n_i(\bm{r})n_k(\bm{r})\right)\lambda_{kljm}p_l(\bm{r})\nonumber\\
&&\times\nabla_m\delta\left({\bf{r}}-{\bf{r}}'\right)\;,
\end{eqnarray}
\begin{eqnarray}
\{n_i({\bf{r}}),g_j({\bf{r'}})\}&=&\left[\nabla_jp_i(\bm{r})\right]\delta\left({\bf{r}}-{\bf{r}}'\right)\nonumber\\
&&-\frac{1}{S}\left(\delta^{T_n}_{ik}(\bm{r})+\frac{Tp_i(\bm{r})p_k(\bm{r})}{S-T}-\frac{Tm_i(\bm{r})m_k(\bm{r})}{S+T}\right)\nonumber\\
&&\times\lambda_{kljm}n_l(\bm{r})\nabla_m\delta\left({\bf{r}}-{\bf{r}}'\right)\;.
\end{eqnarray}

\subsection{Nondissipative Velocities}

For biaxial liquid crystals, the nondissipative velocities associated with the polar and non-polar director fields follow directly from Eqs. (\ref{pg3}) and (\ref{ng3}) and takes the following respective generalized forms~\cite{Brand1982, Brand1981, Jacobsen1981}:
\begin{eqnarray}
V_i^{\bf{p}}&=&-v_j\nabla_jp_i+\lambda^p_{ijk}\nabla_kv_j\;,\\
V_i^{\bf{n}}&=&-v_j\nabla_jn_i+\lambda^n_{ijk}\nabla_kv_j\;,
\end{eqnarray}
where
\begin{eqnarray}
\label{Vp}
\lambda^p_{ijk}&=&\frac12\left[\left(\lambda_1+1\right)p_jm_k+\left(\lambda_1-1\right)p_km_j\right]m_i\nonumber\\
&&-\frac12\large[\left(\lambda_3+1\right)p_jn_k+\left(\lambda_3-1\right)p_kn_j\large]n_i\;,\\
\lambda^n_{ijk}&=&\frac12\left[\left(\lambda_2+1\right)m_jn_k+\left(\lambda_2-1\right)m_kn_j\right]m_i\nonumber\\
&&+\frac12\large[\left(\lambda_3+1\right)p_jn_k+\left(\lambda_3-1\right)p_kn_j\large]p_i\;.
\end{eqnarray}
In Eq. (\ref{Vp}),  only one independent reactive coefficient that is perpendicular to both ${\bf{n}}$ and ${\bf{p}}$ enters into the nondissipative velocity of the polar director field ${\bf{p}}$ due to the orthogonality constraint ${\bf{p\cdot n}}=0$.  As shown in the Appendix, these flow-alignment parameters $\lambda_1$, $\lambda_2$ and $\lambda_3$ can be related to the order parameters and molecular parameters $I$ and $\Delta I$ as follows:
\begin{eqnarray}
\label{lambda1}
\lambda_1&=&\frac{1}{6T}\frac{I}{\Delta I}-\frac12-\frac{S}{3T}\;,\\
\label{lambda2}
\lambda_2&=&\frac{1}{3(S+T)}\frac{I}{\Delta I}-1+\frac{S}{3(S+T)}\;,\\
\lambda_3&=&\frac{1}{3(S-T)}\frac{I}{\Delta I}-1+\frac{S}{3(S-T)}\;.
\label{lambda3}
\end{eqnarray}
In the limit of $T=0$ where the ${\bf{m}}$ and ${\bf{p}}$ directions are no longer distinct and where the symmetry of our mesogenic molecules becomes uniaxial, the flow-alignment parameter $\lambda_1$ drops out of our model, and both $\lambda_2$ and $\lambda_3$ reduce to the expected uniaxial value of $\lambda=\frac13\left(1+\frac{I}{\Delta I}\frac{1}{S}\right)$~\cite{Stark2003}.

For the conserved hydrodynamic variables, the mass-conservation law implies for the nondissipative velocity of mass density
\begin{eqnarray}
V^{\rho}&=&-\nabla_ig_i\;.
\end{eqnarray}
Incorporating the distinction of the three spatial directions, the non-dissipative velocity for the momentum density takes the following form:
\begin{eqnarray}
V^{\bf{g}}_i
&=&-\nabla_j\frac{g_ig_j}{\rho}-\rho\nabla_i\frac{\delta F}{\delta\rho}+\left[\nabla_ip_j\right]\frac{\delta F}{\delta p_j}+\left[\nabla_in_j({\bf{r}})\right]\frac{\delta F}{\delta n_j}\nonumber\\
&&+\nabla_m\left[\frac{1}{2T}\left(\delta^{T_p}_{jk}-\frac{S+T}{S-T}n_jn_k\right)\lambda_{klim}p_l\frac{\delta F}{\delta p_j}\right]\nonumber\\
&&+\nabla_m\left[\frac{1}{S}\left(\delta^{T_n}_{jk}+\frac{Tp_jp_k}{S-T}-\frac{Tm_jm_k}{S+T}\right)\lambda_{klim}n_l\frac{\delta F}{\delta n_j}\right]\nonumber\\
\end{eqnarray}

\subsection{Dissipative Terms}

  Now we turn our attention to the dissipative terms.  The tensor of viscosities will be more complicated due to biaxial symmetry leading to nine independent viscosities~\cite{Liu1981}.  In principle, the lack of ${\bf{p}}\rightarrow -{\bf{p}}$ symmetry would permit new terms to appear in the viscosity tensor.  However, these new terms appear only in terms higher-order in gradients.  Thus, just as in the uniaxial case, the viscosity tensor for the polar biaxial phase would be the same, to the lowest order, as that  for the non-polar counterpart with local $D_{2h}$-symmetry.  Since the form of the viscosity tensor $\eta_{ijkl}$, for which $\sigma^V_{ij}=\eta_{ijkl}U_{kl}$, where $\sigma^V_{ij}$ is the dissipative stress tensor and $U_{kl}=\frac12\left(\nabla_kv_l+\nabla_lv_k\right)$ is the strain-rate tensor, for the biaxial nematics is well known, we shall simply quote the result from literature~\cite{Brand1981, Brand2002}:
%
\begin{eqnarray}
\eta_{ijkl}&=&\alpha_1p_ip_jp_kp_l+\alpha_2n_in_jn_kn_l+\alpha_3\delta_{ij}^{m}\delta_{kl}^{m}\nonumber\\
&&+\alpha_4\left(p_ip_jn_kn_l+p_kp_ln_in_j\right)\nonumber\\
&&+\alpha_5\left(p_ip_j\delta^m_{kl}+p_kp_l\delta^m_{ij}\right)\nonumber\\
&&+\alpha_6\left(n_in_j\delta_{kl}^{m}+n_kn_l\delta_{ij}^{m}\right)\nonumber\\
&&+\alpha_7\left(p_jp_ln_in_k+p_jp_kn_in_l+p_ip_kn_jn_l+p_ip_ln_jn_k\right)\nonumber\\
&&+\alpha_8\left(p_jp_l\delta^m_{ik}+p_jp_k\delta^m_{il}+p_ip_k\delta_{jl}^{m}+p_ip_l\delta^m_{jk}\right)\nonumber\\
&&+\alpha_9\left(n_jn_l\delta^m_{ik}+n_jn_k\delta^m_{il}+n_in_k\delta_{jl}^{m}+n_in_l\delta^m_{jk}\right)\;.\nonumber\\
\end{eqnarray}
%
where
\begin{eqnarray}
\delta^{m}_{ij}&=&\delta_{ij}-p_ip_j-n_in_j\;
\label{Tnp}
\end{eqnarray}
is the operator projecting onto components along ${\bf{m}}={\bf{p}}\times{\bf{n}}$.  Therefore, the dissipative stress tensor $\sigma^V_{ij}$
takes the following form:
\begin{eqnarray}
\sigma^V_{ij}&=&\alpha_1p_ip_jp_kp_lU_{kl}+\alpha_2n_in_jn_kn_lU_{kl}+\alpha_3\delta^m_{ij}U_{kk}\nonumber\\
&&-\alpha_3\delta^m_{ij}p_kp_lU_{kl}-\alpha_3\delta^m_{ij}n_kn_lU_{kl}+\alpha_4p_ip_jn_kn_lU_{kl}\nonumber\\
&&+\alpha_4p_kp_ln_in_jU_{kl}+\alpha_5p_ip_j\delta^m_{kl}U_{kl}+\alpha_5p_kp_l\delta^m_{ij}U_{kl}\nonumber\\
&&+\alpha_6n_in_j\delta^m_{kl}U_{kl}+\alpha_6n_kn_l\delta^m_{ij}U_{kl}+\alpha_7p_jp_ln_in_kU_{kl}\nonumber\\
&&+\alpha_7p_jp_kn_in_lU_{kl}+\alpha_7p_ip_kn_jn_lU_{kl}+\alpha_7p_ip_ln_jn_kU_{kl}\nonumber\\
&&+\alpha_8p_jp_l\delta^m_{ik}U_{kl}+\alpha_8p_jp_k\delta^m_{il}U_{kl}+\alpha_8p_ip_k\delta_{jl}^{m}U_{kl}\nonumber\\
&&+\alpha_8p_ip_l\delta^m_{jk}U_{kl}+\alpha_9n_jn_l\delta^m_{ik}U_{kl}+\alpha_9n_jn_k\delta^m_{il}U_{kl}\nonumber\\
&&+\alpha_9n_in_k\delta_{jl}^{m}U_{kl}+\alpha_9n_in_l\delta^m_{jk}U_{kl}\;.\nonumber\\
\end{eqnarray}

For the polar and nematic director fields ${\bf{p(r)}}$ and ${\bf{n(r)}}$, we must include all dissipative terms allowed by the symmetry.  The odd symmetry of dissipation under time-reversal implies that only dissipative couplings to $\delta F/\delta p_j$ and $\delta F/\delta n_j$ are allowed.  In view of Eq. (\ref{Evo}), the dissipative part of the polar- and nematic-director equations can be schematically written as follows:
\begin{eqnarray}
\frac{dp^D_i}{dt}&=& -\Gamma^{pp}_{ij}\frac{\delta F}{\delta p_j}-\Gamma^{pn}_{ij}\frac{\delta F}{\delta n_j}\;;\\
\frac{dn^D_i}{dt}&=&-\Gamma^{nn}_{ij}\frac{\delta F}{\delta n_j}-\Gamma^{np}_{ij}\frac{\delta F}{\delta p_j}
\end{eqnarray}
where the coefficients $\left(\Gamma^{pp}_{ij}\right)^{-1}$, $\left(\Gamma^{pn}_{ij}\right)^{-1}$, $\left(\Gamma^{nn}_{ij}\right)^{-1}$, $\left(\Gamma^{pn}_{ij}\right)^{-1}$ are the rotational viscosities.  The Onsager principle requires another constraint that $\Gamma^{pn}_{ij}=\Gamma^{np}_{ij}$.   Imposing the unity and orthonormality conditions on the director fields, ${\bf{n}}^2={\bf{p}}^2=1$ and ${\bf{n\cdot p}}=0$, we obtain a total of twelve independent rotational viscosities for the dissipative tensor:
\begin{eqnarray}
\frac{dp^D_i}{dt}&=&-n_i\left[\gamma_1p_j\frac{\delta F}{\delta n_j}+\gamma_2n_j\frac{\delta F}{\delta n_j}+\gamma_3m_j\frac{\delta F}{\delta n_j}\right]\nonumber\\
&&-n_i\left[-\gamma_2p_j\frac{\delta F}{\delta p_j}-\gamma_1n_j\frac{\delta F}{\delta p_j}+\gamma_3m_j\frac{\delta F}{\delta p_j}\right]\nonumber\\
&&-m_i\left[\gamma'_1p_j\frac{\delta F}{\delta n_j}+\gamma'_2n_j\frac{\delta F}{\delta n_j}+\gamma'_3m_j\frac{\delta F}{\delta n_j}\right]\nonumber\\
&&-m_i\left[\gamma_4p_j\frac{\delta F}{\delta p_j}+\gamma_5n_j\frac{\delta F}{\delta p_j}+\gamma_6m_j\frac{\delta F}{\delta p_j}\right]
\label{pD}
\end{eqnarray}
\begin{eqnarray}
\frac{dn^D_i}{dt}&=&-p_i\left[-\gamma_1p_j\frac{\delta F}{\delta n_j}-\gamma_2n_j\frac{\delta F}{\delta n_j}-\gamma_3m_j\frac{\delta F}{\delta n_j}\right]\nonumber\\
&&-p_i\left[\gamma_2p_j\frac{\delta F}{\delta p_j}+\gamma_1n_j\frac{\delta F}{\delta p_j}-\gamma_3m_j\frac{\delta F}{\delta p_j}\right]\nonumber\\
&&-m_i\left[\gamma'_4p_j\frac{\delta F}{\delta n_j}+\gamma'_5n_j\frac{\delta F}{\delta n_j}+\gamma'_6m_j\frac{\delta F}{\delta n_j}\right]\nonumber\\
&&-m_i\left[-\gamma'_2p_j\frac{\delta F}{\delta p_j}-\gamma'_1n_j\frac{\delta F}{\delta p_j}+\gamma'_3m_j\frac{\delta F}{\delta p_j}\right]
\label{nD}
\end{eqnarray}
We can easily see that Eqs. (\ref{pD}) and (\ref{nD}) readily satisfy the constraints that $n_i\left(dn_i/dt\right)=p_i\left(dp_i/dt\right)=0$ and $n_i\left(dp_i/dt\right)+p_i\left(dn_i/dt\right)=0$.  For simplicity, we will set all twelve rotational viscosities to be $\gamma$ in our subsequent calculations.  As in the case of polar unaxial liquid crystals, the only difference between the two sets of equations come from differences in the free energy.

We now collect the reactive and dissipative terms and write down the hydrodynamic equations for the field variables.  We will then consider linearized fluctuations about the ground state of uniform director fields along the ${\bf{n}}$ and ${\bf{p}}$-directions as well as of a uniform density $\rho_0$.  Inertial terms connected with rotations of alignment axes are generally negligible for unaxial systems, and the same will be true for biaxial nematics.  The inertial terms connected with flow should be the same as those for normal liquids, and they can, of course, not be generally neglected even though in many cases they will be small against friction~\cite{Saupe1981}.  Defining the vorticity $\omega_{ij}=\frac12\left(\nabla_iv_j-\nabla_jv_i\right)$ in the director equations, the full hydrodynamic equations are as follows:
\begin{eqnarray}
\frac{\partial\rho}{\partial t}&=&-\bm{\nabla}\cdot{\bf{g(r)}}\;,
\end{eqnarray}
\begin{eqnarray}
\frac{\partial g_i}{\partial t}&=&-\nabla_j\frac{g_ig_j}{\rho}-\rho\nabla_i\frac{\delta F}{\delta\rho}+\left[\nabla_ip_j\right]\frac{\delta F}{\delta p_j}+\left[\nabla_in_j\right]\frac{\delta F}{\delta n_j}\nonumber\\
&&+\nabla_m\left[\frac{1}{2T}\left(\delta^{T_p}_{jk}-\frac{S+T}{S-T}n_jn_k\right)\lambda_{klim}p_l\frac{\delta F}{\delta p_j}\right]\nonumber\\
&&+\nabla_m\left[\frac{1}{S}\left(\delta^{T_n}_{jk}+\frac{Tp_jp_k}{S-T}-\frac{Tm_jm_k}{S+T}\right)\lambda_{klim}n_l\frac{\delta F}{\delta n_j}\right]\nonumber\\
&&+\nabla_j\sigma^V_{ij}\;,\nonumber\\
\end{eqnarray}
\begin{eqnarray}
\frac{\partial p_i}{\partial t}&=&-v_j\nabla_jp_i-\omega_{ij}p_j+\frac{\lambda_1}{2}\left(m_kp_j+m_jp_k\right)U_{kj}m_i\nonumber\\
&&-\frac{\lambda_3}{2}\left(n_kp_j+n_jp_k\right)U_{kj}n_i\nonumber\\
&&-n_i\left[\gamma p_j\frac{\delta F}{\delta n_j}+\gamma n_j\frac{\delta F}{\delta n_j}+\gamma m_j\frac{\delta F}{\delta n_j}\right]\nonumber\\
&&-n_i\left[-\gamma p_j\frac{\delta F}{\delta p_j}-\gamma n_j\frac{\delta F}{\delta p_j}+\gamma m_j\frac{\delta F}{\delta p_j}\right]\nonumber\\
&&-m_i\left[\gamma p_j\frac{\delta F}{\delta n_j}+\gamma n_j\frac{\delta F}{\delta n_j}+\gamma m_j\frac{\delta F}{\delta n_j}\right]\nonumber\\
&&-m_i\left[\gamma p_j\frac{\delta F}{\delta p_j}+\gamma n_j\frac{\delta F}{\delta p_j}+\gamma m_j\frac{\delta F}{\delta p_j}\right]\;,
\end{eqnarray}
\begin{eqnarray}
\frac{\partial n_i}{\partial t}&=&-v_j\nabla_jn_i-\omega_{ij}n_j+\frac{\lambda_2}{2}\left(m_kn_j+m_jn_k\right)U_{kj}m_i\nonumber\\
&&+\frac{\lambda_3}{2}\left(p_kn_j+p_jn_k\right)U_{kj}p_i\nonumber\\
&&-p_i\left[-\gamma p_j\frac{\delta F}{\delta n_j}-\gamma n_j\frac{\delta F}{\delta n_j}-\gamma m_j\frac{\delta F}{\delta n_j}\right]\nonumber\\
&&-p_i\left[\gamma p_j\frac{\delta F}{\delta p_j}+\gamma n_j\frac{\delta F}{\delta p_j}-\gamma m_j\frac{\delta F}{\delta p_j}\right]\nonumber\\
&&-m_i\left[\gamma p_j\frac{\delta F}{\delta n_j}+\gamma n_j\frac{\delta F}{\delta n_j}+\gamma m_j\frac{\delta F}{\delta n_j}\right]\nonumber\\
&&-m_i\left[-\gamma p_j\frac{\delta F}{\delta p_j}-\gamma n_j\frac{\delta F}{\delta p_j}+\gamma m_j\frac{\delta F}{\delta p_j}\right]\;,
\end{eqnarray}
where
\begin{eqnarray}
\frac{\delta F}{\delta \rho}&=&C_1\left(\frac{\delta\rho}{\rho_0^2}\right)-C_2\bm{\nabla}^2\rho+\frac{B}{\rho_0}\nabla_jp_j\;,
\end{eqnarray}
\begin{eqnarray}
\label{Fppp}
\frac{\delta F}{\delta p_j}&=&-\frac{B}{\rho_0}\nabla_j\rho-K_{p1}\nabla_j(\bm{\nabla}\cdot{\bf{p}})+K_{p2}(\bm{\nabla}\times(\bm{\nabla}\times{\bf{p}}))_j\nonumber\\
&&-(K_{p2}-K_{p3})p_s\left(\nabla_sp_k\right)\left(\nabla_jp_k\right)\nonumber\\
&&+(K_{p2}-K_{p3})\nabla_b\left(p_bp_s\left(\nabla_sp_j\right)\right)+O\left(\bm{\nabla}^3\right)\;,
\end{eqnarray}
\begin{eqnarray}
\frac{\delta F}{\delta n_j}&=&-K_{n1}\nabla_j(\bm{\nabla}\cdot{\bf{n}})+K_{n2}(\bm{\nabla}\times(\bm{\nabla}\times{\bf{n}}))_j\nonumber\\
&&-(K_{n2}-K_{n3})n_s\left(\nabla_sn_k\right)\left(\nabla_jn_k\right)\nonumber\\
&&+(K_{n2}-K_{n3})\nabla_b\left(n_bn_s\left(\nabla_sn_j\right)\right)+O\left(\bm{\nabla}^3\right)\;.\nonumber\\
\label{Fnnn}
\end{eqnarray}
The omitted terms in the expressions for $\delta F/\delta\rho$, $\delta F/\delta p_j$, and $\delta F/\delta n_j$ do not contribute to the linear regime of interest.
In spite of the apparent complexity of the biaxial elastic free energy in Eqs. (\ref{Feven}) and (\ref{Fodd}), most of the terms are highly nonlinear and, fortunately, do not come into consideration in the linear hydrodynamics regime.

\section{Hydrodynamic Modes in the Polarized Phase}

In the isotropic phase with full rotational and translational invariance, if we restrict our attention to isothermal processes, the hydrodynamic fields consist only of the four conserved variables:  number density and the three components the momentum density.  The corresponding four hydrodynamic modes are the two propagating sound waves associated with the decay of density and longitudinal momentum fluctuations, as well as two shear modes describing the decay of the two transverse components of the momentum density.  In the less-symmetric unaxial nematic phase, be it macroscopically polar or otherwise, there exists two additional modes associated with the transverse fluctuation of the director field, which arises from the residual unbroken $SO(2)$-symmetry.  The system now possesses six hydrodynamic fields in total, and the modes now separate into a set of two diffusive modes, associated with the velocity and director fluctuations transverse to both the wavevector and equilibrium director orientation, that characterize the coupled decay of vorticity and twist fluctuations~\cite{Kung2006, Stephen1974},  and a set of four remaining longitudinal modes that consist of two propagating sound waves and two diffusive modes describing the coupled decay of splay and velocity, longitudinal to the wavevector, fluctuations~\cite{Stephen1974}.  In a nematic system that exhibits macroscopic polarity, the sound waves receive corrections from the parity-odd terms in the free energy, but the overall structure of the longitudinal modes remains qualitatively the same as the non-polar phase.

We now consider the long wavelength, low frequency dynamics in a uniformly polarized biaxial phase, characterized by uniform equilibrium values $\rho_0$ of the number density, ${\bf{v}}_0=0$ of the flow velocity, ${\bf{p}}_0={\hat{\bf{y}}}$ of the polar director field, and ${\bf{n}}_0={\hat{\bf{z}}}$ of the nematic director field.  In order to evaluate the hydrodynamic modes of the polarized biaxial state, we expand the hydrodynamic equations to linear order in the fluctuations of density, $\delta\rho=\rho-\rho_0$, momentum $\delta{\bf{g}}=\rho_0{\bf{v}}$, polar director field $\delta{\bf{p}}={\bf{p}}-{\hat{\bf{y}}}$, and nematic director field $\delta{\bf{n}}={\bf{n}}-{\hat{\bf{z}}}$ about their equilibrium values.  Since $\bf{p}$ and $\bf{n}$ are orthogonal, there are only three independent variables associated with the broken symmetries, chosen as $\delta n_x$, $\delta n_y$, and $\delta p_x\;(\delta p_z=-\delta n_y)$.  Again, we examine only isothermal processes.  Considering wave-like fluctuations whose space and time dependence is of the form $\exp\left[i{\bf{q\cdot r}}-i\omega t\right]$,  we obtain the linearized equations for the polar biaxial phase:
\begin{widetext}
\begin{eqnarray}
\partial_t\delta\rho_{\bf{q}}&=&-i\rho_0\,\bf{q\cdot v_q}\;,
\label{rhoq222}
\end{eqnarray}
\begin{eqnarray}
\partial_t \delta p^x_{\bf{q}}&=&\delta\rho_{\bf{q}}\left[i\frac{\gamma B}{\rho_0}\left(q_x+q_y+q_z\right)\right]+\delta p^x_{\bf{q}}\,\gamma\left[-K^{(p)}_x({\bf{\hat{q}}})q^2+\Delta K^{(p)}\left(q_xq_z+q_xq_y\right)\right]\nonumber\\
&&+\delta n^{x}_{\bf{q}}\,\gamma\left[-K^{(n)}_x({\bf{\hat{q}}})q^2+\Delta K^{(n)}\left(q_xq_z+q_xq_y\right)\right]\nonumber\\
&&+\delta n^{y}_{\bf{q}}\,\gamma\left[K^{(p)}_z({\bf{\hat{q}}})q^2-K^{(n)}_y({\bf{\hat{q}}})q^2-\Delta K^{(p)}\left(q_xq_z+q_yq_z\right)
+\Delta K^{(n)}\left(q_xq_y+q_yq_z\right)\right]\nonumber\\
&&+v^x_{\bf{q}}\left[i\left(\frac{1+\lambda_1}{2}\right)q_y\right]-v^y_{\bf{q}}\left[i\left(\frac{1-\lambda_1}{2}\right)q_x\right]\;,
\label{pxq222}
\end{eqnarray}
\begin{eqnarray}
\label{nxq111}
\partial_t\delta n^{x}_{\bf{q}}&=&\delta\rho_{\bf{q}}\left[i\frac{\gamma B}{\rho_0}\left(q_x-q_y-q_z\right)\right] +\delta p^x_{\bf{q}}\,\gamma\left[-K^{(p)}_x({\bf{\hat{q}}})q^2-\Delta K^{(p)}\left(q_xq_z+q_xq_y\right)\right]\nonumber\\
&&+\delta n^{x}_{\bf{q}}\,\gamma\left[-K^{(n)}_x({\bf{\hat{q}}})q^2+\Delta K^{(n)}\left(q_xq_z+q_xq_y\right)\right]\nonumber\\
&&+ \delta n^y_{\bf{q}}
\,\gamma\left[-K^{(p)}_z({\bf{\hat{q}}})q^2-K^{(n)}_y({\bf{\hat{q}}})q^2-\Delta K^{(p)}\left(q_xq_z-q_yq_z\right)
+\Delta K^{(n)}\left(q_xq_y+q_yq_z\right)\right]\nonumber\\
&&+v^{x}_{\bf{q}}\left[i\left(\frac{1+\lambda_2}{2}\right)q_z\right] -v^{z}_{\bf{q}}\left[i\left(\frac{1-\lambda_2}{2}\right)q_{x} \right]
\;,
\end{eqnarray}
\begin{eqnarray}
\label{nyq222}
\partial_t\delta n^{y}_{\bf{q}}&=&\delta\rho_{\bf{q}}\left[i\frac{\gamma B}{\rho_0}\left(q_z+q_y-q_x\right)\right] +\delta p^x_{\bf{q}}\, \gamma\left[K^{(p)}_x({\bf{\hat{q}}})q^2+\Delta K^{(p)}\left(q_xq_z+q_xq_y\right)\right]\nonumber\\
&&+\delta n^{x}_{\bf{q}}\, \gamma\left[K^{(n)}_x({\bf{\hat{q}}})q^2-\Delta K^{(n)}\left(q_xq_z+q_xq_y\right)\right]\nonumber\\
&&+\delta n^{y}_{\bf{q}}\, \gamma\left[K^{(p)}_z({\bf{\hat{q}}})q^2+K^{(n)}_y({\bf{\hat{q}}})q^2+\Delta K^{(p)}\left(q_xq_z-q_yq_z\right)-\Delta K^{(n)}\left(q_xq_y+q_yq_z\right)\right]\nonumber\\
&&+ v^{y}_{\bf{q}}\left[i\left(\frac{1+\lambda_3}{2}\right)q_z\right]-v^{z}_{\bf{q}}\left[i\left(\frac{1-\lambda_3}{2}\right)q_{y}\right] \;,
\end{eqnarray}
\begin{eqnarray}
\rho_0\partial_t v^{x}_{\bf{q}}&=&\delta\rho_{\bf{q}}\left[-i\frac{C_1}{\rho_0}q_x-iC_2\rho_0q_{x}q^2+\left(\frac{1+\lambda_1}{2}\right)\left(\frac{B}{\rho_0}\right)q_xq_y\right]\nonumber\\
&&+\delta p^x_{\bf{q}}\large\left[Bq^2_{x}+i\left(\frac{1+\lambda_1}{2}\right)K^{(p)}_x({\bf{\hat{q}}})q^2q_y\large\right]\nonumber\\
&&+\delta n^x_{\bf{q}}\left[i\left(\frac{1+\lambda_2}{2}\right)K^{(n)}_x({\bf{\hat{q}}})q^2q_z\right]\nonumber\\
&&+\delta n^y_{\bf{q}}\left[-Bq_xq_z+i\left(\frac{1+\lambda_1}{2}\right)\Delta K^{(p)}q_xq_yq_z-i\left(\frac{1+\lambda_2}{2}\right)\Delta K^{(n)}q_xq_yq_z \right]\nonumber\\
&&-v^x_{\bf{q}}\left[\alpha_3q_x^2+\alpha_8q_y^2+\alpha_9q_z^2\right]-v^y_{\bf{q}}\left[\left(\alpha_5+\alpha_8\right)q_xq_y\right]-v^z_{\bf{q}}\left[\left(\alpha_6+\alpha_9\right)q_xq_z\right]\;,
\label{vxq222}
\end{eqnarray}
\begin{eqnarray}
\rho_0\partial_tv^{y}_{\bf{q}}&=& \delta\rho_{\bf{q}}\left[-i\frac{C_1}{\rho_0}q_y-iC_2\rho_0q_yq^2+\left(\frac{1-\lambda_3}{2}\right)\left(\frac{B}{\rho_0}\right)q^2_z-\left(\frac{1-\lambda_1}{2}\right)\left(\frac{B}{\rho_0}\right)q^2_x\right]\nonumber\\
&&+\delta p^x_{\bf{q}}\left[Bq_xq_{y}-i\left(\frac{1-\lambda_1}{2}\right)K^{(p)}_x({\bf{\hat{q}}})q^2q_x-i\left(\frac{1-\lambda_3}{2}\right)\Delta K^{(p)}q_xq^2_z\right]\nonumber\\
&&+\delta n^x_{\bf{q}}\left[-i\left(\frac{1+\lambda_3}{2}\right)\Delta K^{(n)}q_xq_yq_z\right]\nonumber\\
&&+\delta n^y_{\bf{q}}\left[-Bq_yq_z-i\left(\frac{1-\lambda_3}{2}\right)K^{(p)}_z({\bf{\hat{q}}})q^2q_z-i\left(\frac{1-\lambda_1}{2}\right)\Delta K^{(p)}q^2_xq_z+i\left(\frac{1+\lambda_3}{2}\right)K^{(n)}_y({\bf{\hat{q}}})q^2q_z\right]\nonumber\\
&&\nonumber\\
&&-v^x_{\bf{q}}\left[\left(\alpha_5+\alpha_8\right)q_xq_y\right]-v^y_{\bf{q}}\left[\alpha_1q_y^2+\alpha_7q^2_z+\alpha_8q^2_x\right]-v^z_{\bf{q}}\left[\left(\alpha_4+\alpha_7\right)q_yq_z\right]\;,
\label{vyq111}
\end{eqnarray}
\begin{eqnarray}
\rho_0\partial_t v^z_{\bf{q}}&=&\delta\rho_{\bf{q}}\left[-i\frac{C_1}{\rho_0}q_z-iC_2\rho_0q_zq^2-\left(\frac{1+\lambda_3}{2}\right)\left(\frac{B}{\rho_0}\right)q_yq_z\right]\nonumber\\
&&+\delta p^x_{\bf{q}}\left[Bq_xq_{z}+i\left(\frac{1+\lambda_3}{2}\right)\Delta K^{(p)}q_xq_yq_z\right]\nonumber\\
&&+\delta n^x_{\bf{q}}\left[-i\left(\frac{1-\lambda_2}{2}\right)K^{(n)}_x({\bf{\hat{q}}})q^2q_x+i\left(\frac{1-\lambda_3}{2}\right)\Delta K^{(n)}q_xq^2_y\right]\nonumber\\
&&+\delta n^y_{\bf{q}}\left[-Bq^2_z+i\left(\frac{1+\lambda_3}{2}\right)K^{(p)}_z({\bf{\hat{q}}})q^2q_y+i\left(\frac{1-\lambda_2}{2}\right)\Delta K^{(n)}q^2_xq_y-i\left(\frac{1-\lambda_3}{2}\right)K^{(n)}_y({\bf{\hat{q}}})q^2q_y\right]\nonumber\\
&&-v^x_{\bf{q}}\left[\left(\alpha_6+\alpha_9\right)q_xq_z\right]-v^y_{\bf{q}}\left[\left(\alpha_4+\alpha_7\right)q_yq_z\right]-v^z_{\bf{q}}\left[\alpha_2q^2_z+\alpha_9q^2_x+\alpha_7q^2_y\right]\;,
\label{vzq}
\end{eqnarray}
\end{widetext}
where we have defined the following quantities:
\begin{eqnarray}
K^{(n)}_x({\bf{\hat{q}}})&=&\left[K_{n1}q_x^2+ K_{n2}q^2_y+K_{n3}q^2_z\right]/q^2\;,
\end{eqnarray}
\begin{eqnarray}
K^{(n)}_y({\bf{\hat{q}}})&=&\left[K_{n1}q_y^2+ K_{n2}q^2_x+K_{n3}q^2_z\right]/q^2\;,
\end{eqnarray}
\begin{eqnarray}
K^{(p)}_x({\bf{\hat{q}}})&=&\left[K_{p1}q^2_x+K_{p2}\,q^2_z+K_{p3}\,q^2_y\right]/q^2\;,\\
K^{(p)}_z({\bf{\hat{q}}})&=&\left[K_{p1}q^2_z+K_{p2}\,q^2_x+K_{p3}\,q^2_y\right]/q^2\;,
\end{eqnarray}
\begin{eqnarray}
\Delta K^{(p)}&=&K_{p2}-K_{p1}\;,\\
\Delta K^{(n)}&=&K_{n2}-K_{n1}\;,
\end{eqnarray}
The system has seven hydrodynamic modes described by dispersion relations $\imath\omega({\bf{q}})$.  Stable modes describing decaying fluctuations correspond to ${\Re}[\imath\,\omega({\bf{q}})]>0$, while a negative sign in $\Re[\imath\omega({\bf{q}})]$ signals the instability of the corresponding uniform state.  The seven hydrodynamic modes are all coupled together.  There is
no longer a separation between sets of {\it{transverse}} and
{\it{longitudinal}} modes.  In terms of the basis of field
variables $(\delta\rho_{\bf{q}},\, \delta p^x_{\bf{q}},\, \delta
n^x_{\bf{q}},\, \delta n^y_{\bf{q}},\, v^x_{\bf{q}},\,
v^y_{\bf{q}},\, v^z_{\bf{q}})$, probing their mode structures
amount to diagonalizing a linear system of seven coupled
differential equations.  While the numerically obtained
expressions of the normal modes and their associated
eigenfrequencies, expanded to $O(q^2)$, are rather complicated,
the salient feature of the normal modes in the polar biaxial phase
is that the new parity-odd terms, proportional to the elastic
constant $B$, give quantitative corrections to the sound modes as
well as to the diffusive modes.  In total, there are a pair of
sound modes with the usual (when compared with other fluids) sound
speed of $\pm \sqrt{C_1/\rho_0}$ and a complicated anisotropic
decay rate, and five modes with dispersion of order $O(q^2)$.
Comparing with the hydrodynamic modes found for unaxial liquid
crystals, there always exists a new purely diffusive mode which
may be identified as the Goldstone mode of the uniaxial-biaxial
transition~\cite{Liu1981}.  Overall, the diffusion peaks contain
angular dependence as all three spatial axes are now identified in
the polarized biaxial phase.  In particular, we will express our
results in terms of spherical coordinates:
${\hat{q}}_x=\sin\theta\cos\phi$,
${\hat{q}}_y=\sin\theta\sin\phi$, and ${\hat{q}}_z=\cos\theta$.
Expanding the hydrodynamic frequencies $\omega$ to quadratic order
$(i\omega=\eta_1q+\eta_2q^2)$, and solving the eigenvalue problem
perturbatively, we obtain the following explicit expressions for
the propagating sound modes:
\begin{eqnarray}
i\omega^s_{\pm}&\approx&\pm i\,q\sqrt{\frac{C_1}{\rho_0}}-q^2\mathcal{D}\left(\phi,\theta\right)\;,
\end{eqnarray}
where
\begin{widetext}
\begin{eqnarray}
\mathcal{D}\left(\phi,\theta\right)&=&\left.\frac{\gamma  B^2}{8C_1}\left[-1-3\cos 2\theta+2\sin^2\theta\left(\cos 2\phi+\sin2\phi\right)+2\sin 2\theta\left(2\cos\phi-\sin\phi\right)\right]\right.\nonumber\\
&&+\nu_2\cos^4\theta+\sin^4\theta\big[\nu_3\cos^4\phi+\nu_1\sin^4\phi+2(\nu_5+2\nu_8)\cos^2\phi\sin^2\phi\big]\nonumber\\
&&+\cos^2\theta\sin^2\theta\big[2(\nu_6+2\nu_9)\cos^2\phi+(\nu_4+3\nu_7)\sin^2\phi+(\nu_4+\nu_7)\cos\phi\sin\phi\big]\;,
\label{firstsounds}
\end{eqnarray}
\end{widetext}
where we have defined the kinematic viscosities as $\nu_i=\alpha_i/2\rho_0$ as they are the natural quantities that determine momentum relaxation.  In particular, along the three specified directions along $q_x$ where $(\phi, \theta)=(0,\pi/2)$, along $q_y$ where $(\phi, \theta)=(\pi/2,\pi/2)$, and along $q_z$ where $\theta=0$, the diffusive components of the first sound mode reduce to the following limits:
\begin{eqnarray}
&&\mathcal{D}\left(\phi=0,\theta=\frac{\pi}{2}\right)=\left(\frac{\gamma B^2}{2C_1}+\nu_3\right)q_x^2\\
&&\mathcal{D}\left(\phi=\frac{\pi}{2},\theta=\frac{\pi}{2}\right)=\nu_1q_y^2\\
&&\mathcal{D}\left(\phi,\theta=0\right)=\left(-\frac{\gamma B^2}{2C_1}+\nu_2\right)q^2_z
\end{eqnarray}
The sound modes are anisotropic, as expected from the calculation
of the structure factor and the free-energy analysis in section
II. In addition, the coupling of density to splay fluctuation can
render the sound modes unstable for certain direction of
propagation, although
viscous dissipation tends to stabilize them.  Assuming, for
simplicity, that $\nu_i=\nu$ for all values of $i$, we find that the
fastest growth of fluctuations occurs for ${\bf q}$ in the $xz$-plane, corresponding to $q_y=0$ or $\phi=0$. In this case we
obtain
\begin{equation} \label{soundinstab}
\mathcal{D}\left(\phi=0,\theta\right)/\nu=1+\sin^2
2\theta+\alpha\big(\sin 2\theta-\cos 2 \theta\big)\;,
\end{equation}
where $\alpha=\gamma B/(2\nu C_1)$. If
$\mathcal{D}\left(\phi,\theta\right)<0$ the modes are unstable.
This occurs provided $\alpha>1$. For small values of $\alpha-1$
the sound waves are unstable for all angles $\theta<\theta_c(\alpha)$,
with $\theta_c\approx(1+\alpha)/2$. 

The remaining five modes all have dispersion relations that are at
least of order $O(q^2)$. The corresponding normal frequencies can
be obtained from the solutions of a fifth-order equation. Their
exact expressions are quite complicated, and we will specialize to
a few limiting cases below whose physical properties can be
readily illuminated.

In the incompressible limit, the sounds modes will have leading contribution from $O(q^2)$ just as what we found for the polar unaxial case, as terms proportional to $B$ (and $D$ to higher orders) vanish.  The vanishing of these partiy-odd terms can be readily seen in the linearized regime in Eq. (\ref{firstsounds}) of which the $q^2$ terms contain contributions quadratic in $B$.  The physical nature of this leading $O(q^2)$-contribution will depend on the relative magnitude of the various viscous coefficients.  To illustrate this point concretely, we now consider the one-constant approximation, often employed in numerical simulations of liquid-crystal systems, in which the corresponding mode structure simplifies considerably.  In this approximation, we neglect the anisotropy in the elastic constants along both polar $(\bm{p})$ and non-polar $(\bm{n})$ directions, setting $K_{p1}=K_{p2}=K_{p3}=K_p$ and $K_{n1}=K_{n2}=K_{n3}=K_n$.  We will consider, in the subsequent subsections, firstly the three limiting cases in which we set $K_p=0$ while keeping a nonvanishing $K_n$ with linearized fluctuations along only $q=q_x$, $q=q_y$ or $q=q_z$, separately.  We will then consider three additional limiting cases,  where fluctuations are again only found along each of the three distinct directions but with $K_n=0$ and a nonvanishing $K_p$.

\subsection{$K_p=0$}

When we set $K_p=0$, the mode structure becomes especially simple when we consider only fluctuations along $q_y$.  In this limit, we obtain the following expression for the quintic algebraic equation that governs the $O(q^2)$-contributions to the eigenmodes $(i\omega=\eta_1q+\eta_2q^2)$:
\begin{widetext}
\begin{eqnarray}
\eta_2\left(\eta_2-2\nu_8\right)\left[C_1\eta_2^3+2\nu_7C_1\eta_2^2+\frac{C_1K_n}{4\rho_0}\left(1-\lambda_3\right)^2\eta_2+\frac{\gamma C_1K_n^2}{4\rho_0}\left(1-\lambda_3\right)^2\right]&=&0
\label{Kp0qy}
\end{eqnarray}
\end{widetext}
As expressed in Eq. (\ref{Kp0qy}), we readily see that there are two modes that decouple from the rest.  The first is the diffusive mode described by the dispersion relation $i\omega=2q^2\nu_8$ that controls the decay of $v_x$, which decouples in this limit.  The second mode  has dispersion relation $i\omega=0$ and describes the relaxation of $\delta p_x$, which does not decay since we set $K_p=0$.  The remaining three modes are governed by the solution of the cubic equation in square brackets in Eq. (\ref{Kp0qy}).  Using Mathematica, we see that they consist of a pair of complex conjugate solutions corresponding to propagating modes with speed of propagation of $O(q^2)$. The third solution is real, corresponding to a diffusive mode.  The nature of these solutions, and hence the stability of the three corresponding modes can be examined using the Descartes' rule of signs from algebraic theory:  since all four coefficients in the cubic equation have the same positive sign ($\nu_7>0$),  there are no positive real zeroes to the cubic equation.  Likewise, when we substitute $\eta_2\rightarrow -\eta_2$ in the cubic equation, we observe three sign-changes between the four terms, when written in descending algebraic power, and the Descartes' rule of signs indicates that there would consequently be either three or one real negative roots to the cubic equation.  Since we already know that two of the three roots are in general complex conjugates, the remaining one real root must hence in general be negative and, thus, corresponds to a physically stable eigenmode. The stability of the propagating modes, on the other hand, is determined by the proper sign of their real part, and an analysis of its precise nature can only be preformed with the knowledge of its full expression.

The mode structure for uni-directional fluctuations along the other two directions, along $q=q_x$ and $q=q_z$, are both considerably more complicated, as the equations governing the $O(q^2)$-contributions to their respective eigenmodes are full quintic equations.  Moreover, the nature of the algebraic roots and the corresponding stability of the eigenmodes would, in general, depend upon the relative magnitudes of the various elastic and viscosity coefficients.  The detailed elucidation of the general hydrodynamic mode structure of the biaxial phases, formed by bent-core nematogens, will be presented elsewhere~\cite{Kung2007}.

\subsection{$K_n=0$}

Analogously, we can consider the simplified limit of nonvanishing $K_p$ and set $K_n=0$.  In the case of fluctuations occuring only along $q=q_x$, we also find a complicated mode structure governed by the following factorized quintic equation:
\begin{widetext}
\begin{eqnarray}
\eta_2\left(\eta_2-2\nu_9\right)\Big\{-C_1\eta_2^3+\left(B^2\gamma-2\nu_8C_1\right)\eta_2^2&&-\left[\left(2\gamma^2K_p-\frac{\left(1-\lambda_1\right)^2}{4\rho_0}\right)\left(B^2-C_1K_p\right)-2\gamma B^2\nu_8\right]\eta_2\nonumber\\
&&\left.+\left(4\nu_8\gamma^2K_p-\frac{\left(1-\lambda_1\right)^2}{4\rho_0}\right)\left(B^2-C_1K_p\right)K_p\gamma\right\}
=0
\label{Kn}
\end{eqnarray}
\end{widetext}
In this limit, $v_z$ is decoupled, and its decay is described by a diffusive mode $i\omega=2q^2\nu_9$.  There is a mode $i\omega=0$ corresponding to the decay of $\delta n_x$, which does not decay since we set $K_n=0$.  Of the remaining three modes given by the solutions of the cubic equation contained in the curly brackets of Eq. (\ref{Kn}), two are complex conjugates and one is real. The stability of the real mode is controlled by the relative magnitude of the various viscosity and elastic coefficients, one of which includes the factor of $\left(B^2-K_pC_1\right)$.  Comparing our free energy analysis in Section II and the examination of the eigenmode equation in Eq. (\ref{Kn}), we find that the condition signaling the onset of splay instability as found in Section II, namely, $\vert B\vert > \sqrt{K_{p}C_1}$ may actually be a necessary but not sufficient requirement for mode instability.  Its rigorous confirmation would require the full analysis of the eigenvalue problem~\cite{Kung2007}.

The mode structures for the other two limits, in which fluctuations are found along only the directions of $q=q_y$ or $q=q_z$, are again governed by the full quintic algebraic equation.  Needless to say, the mode expressions become even more cumbersome when we re-introduce anisotropy in the elastic constants, though they all share several overall features.  In general, the precise physical nature of all five modes, whether diffusive or propagating, depends on the relative magnitude of the various viscosity coefficients and elastic constants.  When the solutions to the full quintic equation, and thus the corresponding hydrodynamic modes, are complex conjugates,  they signify that the leading contribution to their propagating components are of order $O(q^2)$.  

\section{Conclusion}

In conclusion, we have examined in this work the hydrodynamics of a biaxial nematic system which is macroscopically polar, with local $C_{2v}$-symmetry.  Our work is motivated by the recent experimental discovery of the long-sought biaxial nematic phase composed of microscopically polar bent-core mesogens.  Our present results are extension to those previously obtained for the hydrodynamics of polar uniaxial liquid crystals.  In spite of the much more complicated structure of the hydrodynamic equations due to the complete breaking of rotational invariance and to the discrete orthorhombic symmetry, we have found that polarity and nematic biaxiality can coexist, although the system exhibits an  intrinsic bulk splay instability for a range of parameters and along certain directions of the wavevector.  Moreover, we have found that the newly allowed parity-odd terms in the free energy as well as the orthonormality constraints between the directors in the biaxial system provide additional corrections to the hydrodynamic mode expressions and render them highly anisotropic.  Polarity does not, however, alter the linear hydrodynamics of the biaxial phase in the incompressible limit, just as in the more symmetrical uniaxial state.  In deriving the hydrodynamic equations using the Poisson-bracket formalism, we have also computed the flow-alignment parameters along the three symmetry axes of the polar phase in terms of microscopic parameters associated with the molecular geometry of the constituent biaxial mesogens.

It is our hope that the results presented in this paper will contribute to the theoretical understanding of biaxial nematic phases, which have continually generated great interest from the scientific community due to their many fascinating properties as well as their immense potential for technological applications.  In particular, there has been a recent focus on the phenomenon of flexoelectricity~\cite{Harden2006}--an effect uniquely enhanced by the particular boomerang shape of these liquid crystals--that carries the potential of serving as a basis for environmentally friendly micro-power generators.  Furthermore, our study of the hydrodynamic modes would complement the experimental study of the dynamics of the phases of these mesogenic systems using techniques such as dynamic light scattering~\cite{Stojadinovic2002}.   It is our further hope that our results, presented here in the context of equilibrium physics, will also serve as a starting point in understanding the behavior of nonequilibrium systems where polarity arises from actively driven mechanisms.

\begin{acknowledgments}
W.K. and M.C.M. acknowledge support by NSF Grants DMR-0219292 and DMR-0305497.
\end{acknowledgments}

\appendix

\section{Derivation of Flow-Alignment Parameters}

In this Appendix, we present details on the derivation of our flow-alignment parameter expressions found in Eqs (\ref{lambda1}), (\ref{lambda2}), and (\ref{lambda3}), defined via the following:
\begin{eqnarray}
\label{VpA}
\lambda^p_{ijk}&=&\frac12\left[\left(\lambda_1+1\right)p_jm_k+\left(\lambda_1-1\right)p_km_j\right]m_i\nonumber\\
&&-\frac12\large[\left(\lambda_3+1\right)p_jn_k+\left(\lambda_3-1\right)p_kn_j\large]n_i\;,\\
%
\lambda^n_{ijk}&=&\frac12\left[\left(\lambda_2+1\right)m_jn_k+\left(\lambda_2-1\right)m_kn_j\right]m_i\nonumber\\
&&+\frac12\large[\left(\lambda_3+1\right)p_jn_k+\left(\lambda_3-1\right)p_kn_j\large]p_i\;,
\label{VnA}
\end{eqnarray}
and
\begin{eqnarray}
\label{lambda1A}
\lambda_1&=&\frac{1}{6T}\frac{I}{\Delta I}-\frac12-\frac{S}{3T}\;,\\
\label{lambda2A}
\lambda_2&=&\frac{1}{3(S+T)}\frac{I}{\Delta I}-1+\frac{S}{3(S+T)}\;,\\
\lambda_3&=&\frac{1}{3(S-T)}\frac{I}{\Delta I}-1+\frac{S}{3(S-T)}\;.
\label{lambda3A}
\end{eqnarray}
Starting from Eqs. (\ref{pg3}) and (\ref{ng3}),
\begin{eqnarray}
\label{pgA}
\{p_i({\bf{r}}),g_j({\bf{r'}})\}&=&\frac{1}{2T}\left[\delta^{T_p}_{ik}-\left(\frac{S+T}{S-T}\right)n_in_k\right]\nonumber\\
&&\times\left\{Q_{kl}({\bf{r}}),g_j({\bf{r'}})\right\}p_l({\bf{r}})\;,\\
\{n_i({\bf{r}}),g_j({\bf{r'}})\}&=&\frac{1}{S} \left[\delta^{T_n}_{ik}+\frac{T}{S-T}p_ip_k-\frac{T}{S+T}m_im_k\right]\nonumber\\
&&\times\left\{Q_{kl}({\bf{r}}),g_j({\bf{r'}})\right\}n_l({\bf{r}})\;,\nonumber\\
 \label{ngA}
\end{eqnarray}
and applying Eq. (\ref{V}),
\begin{eqnarray}
V_a({\bf{r}})&=&-\int d^3r'\,\left\{\Phi_a({\bf{r}}), \Phi_b({\bf{r'}})\right\}\frac{\partial\mathcal{H}}{\partial\Phi_b({\bf{r'}})}\;,
\label{VA}
\end{eqnarray}
to the director fields ${\bf{p(r)}}$ and ${\bf{n(r)}}$, we obtain
\begin{eqnarray}
\label{VpA}
V_i^{\bf{p}}&=&-v_j\nabla_jp_i\nonumber\\
&&+\left[\left(\frac{m_im_k}{2T}-\frac{n_in_k}{S-T}\right)\lambda_{kjlm}p_j\right]\left(\nabla_mv_l\right)\;,\nonumber\\
\\
V_i^{\bf{n}}&=&-v_j\nabla_jn_i\nonumber\\
&&+\left[\left(\frac{m_im_k}{S+T}+\frac{p_ip_k}{S-T}\right)\lambda_{kjlm}n_j\right]\left(\nabla_mv_l\right)\;.\nonumber\\
\label{VnA}
\end{eqnarray}
Using Eq. (\ref{lambdaijklA}) and the biaxial form of the alignment tensor in Eq. (\ref{QdefA})
\begin{eqnarray}
\lambda_{ijkl}&=&\delta_{ik}Q_{jl}+\delta_{jk}Q_{il}-\frac23\delta_{ij}Q_{kl}-\frac23\delta_{kl}Q_{ij}\nonumber\\
&&+\frac{I}{6\Delta I}\left(\delta_{ik}\delta_{jl}+\delta_{il}\delta_{jk}-\frac23\delta_{ij}\delta_{kl}\right)\nonumber\\
&&+O(Q^2_{ij})\;,
\label{lambdaijklA}
\end{eqnarray}
\begin{eqnarray}
Q_{ij}({\bf{r}})
&=&S({\bf{r}})\left[n_i({\bf{r}})n_j({\bf{r}})-\frac13\delta_{ij}\right]\nonumber\\
&&+T({\bf{r}})\left[p_i({\bf{r}})p_j({\bf{r}})-m_i({\bf{r}})m_j({\bf{r}})\right]\;,\nonumber\\
\label{QdefA}
\end{eqnarray}
we obtain the following, upon further simplifying:
\begin{widetext}
\begin{eqnarray}
\label{1}
n_in_k\lambda_{kjlm}p_j&=&\left[\left(\frac{I}{6\Delta I}+T-\frac{S}{3}\right)n_lp_m+\left(\frac{I}{6\Delta I}+\frac23S\right)n_mp_l\right]n_i\;,\\
\label{2}
m_im_k\lambda_{kjlm}p_j&=&\left[\left(\frac{I}{6\Delta I}+T-\frac{S}{3}\right)m_lp_m+\left(\frac{I}{6\Delta I}-T-\frac{S}{3}\right)m_mp_l\right]m_i\;,\\
\label{3}
m_im_k\lambda_{kjlm}n_j&=&\left[\left(\frac{I}{6\Delta I}+\frac23 S\right)m_lm_m+\left(\frac{I}{6\Delta I}-T-\frac{S}{3}\right)m_mn_l\right]m_i\;,\\
p_ip_k\lambda_{kjlm}n_j&=&\left[\left(\frac{I}{6\Delta I}+\frac23S\right)p_ln_m+\left(\frac{I}{6\Delta I}+T-\frac{S}{3}\right)p_mn_l\right]p_i\;.
\label{4}
\end{eqnarray}
\end{widetext}
Further decomposing Eqs. (\ref{1})-(\ref{4}) into symmetric and antisymmetric parts and comparing with Eqs. (\ref{VpA}) and (\ref{VnA}), we obtain the desired results in Eqs. (\ref{lambda1A})-(\ref{lambda3A}).


\begin{references}
\bibitem{Footnote}  In the limit of $\psi=\pi/2$, we have a linear molecule of three masses.  Picking the origin of the coordinate system at the location of the central mass $M$, the components of the moment-of-inertia tensor are $I_{11}=2mL^2=I_{22}$, $I_{33}=0$ and $I=4mL^2$.
\end{references}
\end{document}